\documentclass[sigconf]{acmart}

\AtBeginDocument{%
  \providecommand\BibTeX{{%
    \normalfont B\kern-0.5em{\scshape i\kern-0.25em b}\kern-0.8em\TeX}}}

\usepackage{graphics}
\usepackage{xspace}
\usepackage{soul}
\usepackage{siunitx}
\usepackage{enumitem}
\usepackage{balance}
\usepackage{graphics}
\usepackage{xspace}
\usepackage{color, colortbl, xcolor}
\usepackage{tikz}
\usepackage{subcaption}
\usepackage{fixltx2e}
\usepackage{bm}
\usepackage{makecell}

\def\S{Sec.\xspace}

\def\insitu{\textit{in situ}\xspace}
\def\ie{\textit{i.e.,}\xspace}
\def\etal{\textit{et al.}\xspace}

\def\eg{\textit{e.g.,}\xspace}
\def\incl{\textit{incl.}\xspace}

\copyrightyear{2023} 
\acmYear{2023} 
\setcopyright{rightsretained} 
\acmConference[DIS Companion '23]{Designing Interactive Systems Conference}{July 10--14, 2023}{Pittsburgh, PA, USA}
\acmBooktitle{Designing Interactive Systems Conference (DIS Companion '23), July 10--14, 2023, Pittsburgh, PA, USA}
\acmDOI{10.1145/3563703.3596640}
\acmISBN{978-1-4503-9898-5/23/07}

\author{Chen Chen}
\email{chenchen@ucsd.edu}
\orcid{0000-0001-7179-0861}
\affiliation{%
  \department{Computer Science and Engineering}
  \institution{University of California San Diego}
  \city{La Jolla}
  \state{CA}
  \country{United States}
}

\author{Ella T. Lifset}
\email{etlifset@ucsd.edu}
\orcid{0000-0003-2640-3206}
\affiliation{%
  \department{Biological Sciences}
  \institution{University of California San Diego}
  \city{La Jolla}
  \state{CA}
  \country{United States}
}

\author{Yichen Han}
\authornote{The author contributed to the project while at the University of California San Diego.}
\email{yichenha@andrew.cmu.edu}
\orcid{0000-0002-2997-2140}
\affiliation{%
  \department{Electrical and Computer Engineering}
  \institution{Carnegie Mellon University}
  \city{Pittsburgh}
  \state{PA}
  \country{United States}
}

\author{Arkajyoti Roy}
\email{aroy@ucsd.edu}
\orcid{0000-0003-0597-6582}
\affiliation{%
  \department{Department of Mathematics}
  \institution{University of California San Diego}
  \city{La Jolla}
  \state{CA}
  \country{United States}
}

\author{Michael Hogarth}
\email{mihogarth@ucsd.edu}
\orcid{0000-0002-4264-1258}
\affiliation{%
  \department{School of Medicine}
  \institution{University of California San Diego}
  \city{La Jolla}
  \state{CA}
  \country{United States}
}

\author{Alison A. Moore}
\email{alm123@ucsd.edu}
\orcid{0000-0003-2989-4346}
\affiliation{%
  \department{School of Medicine}
  \institution{University of California San Diego}
  \city{La Jolla}
  \state{CA}
  \country{United States}
}

\author{Emilia Farcas}
\email{efarcas@ucsd.edu}
\orcid{0000-0001-6485-0141}
\affiliation{%
  \department{Qualcomm Institute}
  \institution{University of California San Diego}
  \city{La Jolla}
  \state{CA}
  \country{United States}
}

\author{Nadir Weibel}
\email{weibel@ucsd.edu}
\orcid{0000-0002-3457-4227}
\affiliation{%
  \department{Computer Science and Engineering}
  \institution{University of California San Diego}
  \city{La Jolla}
  \state{CA}
  \country{United States}
}

\renewcommand{\shortauthors}{Chen~\etal}



\begin{document}

\title[How do Older Adults Set Up Voice Assistants?]{How do Older Adults Set Up Voice Assistants? Lessons Learned from a Deployment Experience for Older Adults to Set Up Standalone Voice Assistants}

\begin{abstract}
While standalone Voice Assistants (VAs) are promising to support older adults’ daily routine and wellbeing management, onboarding and setting up these devices can be challenging. Although some older adults choose to seek assistance from technicians and adult children, easy set up processes that facilitate independent use are still critical, especially for those who do not have access to external resources. We aim to understand the older adults' experience while setting up commercially available voice-only and voice-first screen-based VAs. Rooted in participants observations and semi-structured interviews, we designed a within-subject study with $10$ older adults using Amazon Echo Dot and Echo Show. We identified the values of the built-in touchscreen and the instruction documents, as well as the impact of form factors, and outline important directions to support older adult independence with VAs.
\end{abstract}

\begin{teaserfigure}
    \centering
    \includegraphics[width=\textwidth]{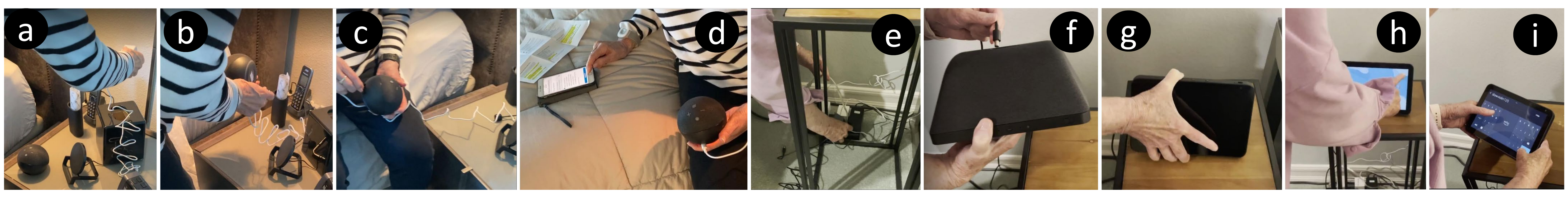}
    \Description[Example demonstrative figures of how to setup Echo Dot and Echo Show.]{(a) The demonstrative older adult is attempting to plug the power plug of the Echo Dot into the outlet behind the closet. (b) The demonstrative older adult is attempting to connect the cable with the Echo Dot. (c) The demonstrative older adult is sitting on her bed and attempting to enable the setup mode by holding the setup button on the top of the Echo Dot. (d) The demonstrative older adults is attempting to finishing the set up workflow with the mobile application. (e) The demonstrative older adult is attempting to plug the power plug of the Echo Show into the outlet underneath her desk. (f) The demonstrative older adult is trying to connect the power cable with the Echo Show. (g) The demonstrative older adult is placing the Echo Show on the desk. (h) The demonstrative older adult is trying to connect the Echo Show with her home's WiFi by clicking the touchscreen, (j) The demonstrative older adult is inputting the WiFi credentials on the Echo Show.}
    \caption{We aim to understand older adults' experience on setting up standalone Voice Assistants (VAs). During the study, participants were instructed to set up \href{https://www.amazon.com/Echo-Dot/dp/B07XJ8C8F5}{Amazon Echo Dot} (a - d) and \href{https://www.amazon.com/dp/B084DCJKSL}{Echo Show} (e - i) --- two mainstream VAs with and without built-in displays.}
    \label{fig::teaser}
\end{teaserfigure}

\maketitle

\section{Introduction}
The needs of ``aging in place'' and living independently, instead of in  assisted living facilities, have been widely recognized in existing research~\cite{Eckert2004, Matsumoto2016, Boldy2011}.
While interacting with digital information is becoming increasingly important, ubiquitously accessing them can be challenging for aging populations~\cite{Marting1997}.
With hands-free and eyes-free interaction modalities to handle the requests, voice user interfaces and conversational Voice Assistants~(VAs) (\eg~Alexa) are promising to support older adults' daily routine and wellbeing management~\cite{Chen2021assets}.  
While recent voice-first screen-based VAs (\eg~Echo Show) outlined new ways to deliver richer information and interactions through a built-in touchscreen~\cite{Whitenton2017}, setting up and maintaining such devices~(\eg~charging, upgrading and rebooting them) might still be challenging for older adults, which might hinder the adoption of the technology~\cite{Heinz2013}.

Standalone VAs (\eg~Echo Speakers) are expected to be permanently attached to the environment. Due to the fact that they only need to be set up once, they are promising devices in terms of their reduced need for maintenance, which allows users to use them more ubiquitously.
However, in order to get through the first step --- \ie setting up these devices --- older adults often ask for technical support from technicians and family members~\cite{Pang2021}. Among the most common reason is the need for help to finish "last-mile" tasks (\eg~connecting VAs to the home's WiFi).
Independent use and setup of technology is especially critical when social distancing or isolation is necessary --- such as  in the case of the COVID-19 pandemic --- and when technology support people are not available.
Although a wide variety of research has explored the user experience of VAs for older adults~\cite{Chen2021assets, Trajkova2020}, no research work looked at the critical onboarding phase, when these devices need to be set up by the older adult.
While few prior works (\eg~\cite{Pang2021}) have identified specific challenges for older adults when setting up general technologies~(\eg~phone and laptop), the insights regarding setting up VAs in particular are still unknown.

We present a {\it preliminary} qualitative study that aims to understand older adults' experience while setting up commercially-available standalone voice-only and voice-first screen-based VAs~\cite{Bajorek2018}. 
We used Amazon \href{https://www.amazon.com/Echo-Dot/dp/B07XJ8C8F5}{Echo Dot} and \href{https://www.amazon.com/dp/B084DCJKSL}{Echo Show} as the devices for our study due to their dominant market share~\cite{Kinsella2020}.
Rooted in participant observation and semi-structured interviews, we designed an in-person \mbox{within-subject study} and reported our findings from $10$~older adults, with an average age of $79.20$.
%
Overall, our results outline three key lessons learned and identify specific affordances of the touchscreen and the instruction documents, as well as the impact of form factors. 

\section{Related Work}\label{sec::related}
Many prior research has investigated the general user experience of standalone VAs from an older adults' perspective.
For example, Trajkova~\etal~\cite{Trajkova2020} found that the long-term adoptions of VAs for general uses depends on older adults' abilities and beliefs, which was evidenced by the fact that most participants become non-users due to the lack of perceived usefulness.
Other research, \eg~\cite{Chen2021assets, Shade2021, Charles2021, Lifset2020, Han2022, Chen2021, Mrini2021}, outlined the potential usefulness of using standalone VAs for well-being management, yet today's VAs can hardly support it. 
While those works pointed to future opportunities, Nallam~\etal~\cite{Nallam2020} suggested that the adoption of such technologies could be affected by barriers such as confidentiality risks and receiving trusted information.
In addition, \mbox{Jaonsons~\etal~\cite{Jansons2022}} implied that the low interactivity of current devices might not completely support the older adults' needs when it comes to managing their wellbeing.
Similarly, Choi~\etal~\cite{Choi2018} indicated that the most frequent use cases are querying practical questions and managing tasks, not necessarily health and wellbeing.
Kim~\etal~\cite{Sunyoung2021} identified challenges involving the unfamiliarity with VAs as well as the functional errors of the limited speech technologies.

These works explored the use of VAs for general, wellbeing, or healthcare-related use cases, assuming that the technology is readily usable by older adults. However, one of the key access barriers to using VAs is to be able to properly set up them. We focus on the device's setup and onboarding phases that are indispensable for heterogeneous subsequent long-term uses. 
During this phase, many older adults choose to seek assistance from technical services, their neighbors, or family member, however we believe that setting up devices {\it independently} and being able to successfully onboard are critical steps that need to be supported, especially for those who do not have access to such technology support resources~\cite{Pang2021}.
To address this, our study takes a deeper look at how existing commercially-available, voice-only, and voice-first screen-based standalone VAs currently support (or do not support) such {\it independent} device setup and the onboarding process.

\section{Methods}
We recruited $10$ older adult participants (P1 - P10, \incl~six females and four males, see Table~\ref{fig::participant}), with the ages range from $70$ to $97$ ($M = 79.20$, $SD = 7.70$).
Participants were recruited from the \href{https://health.ucsd.edu}{UC San Diego Health} and the \href{https://www.viliving.com/}{Vi at La Jolla Village}.
All participants were able to live independently in either a standalone home or a retirement community.
We also reported the level of education and their occupation before retirement, which might be correlated with their prior exposure on computing devices that might affect their abilities on setting up designated VAs.
While two participants {\it disagreed}, five and three participants {\it agreed} and {\it strongly agreed} that they were familiar with VAs, only one participant (P2) had previous experience setting up an Echo Dot, and no participants had set up the Echo Show before the study.
Our study was conducted in-person in participants' residences. 
Our work has been approved by the Institutional Review Board~(IRB).

\begin{table*}[t]
\centering
\begin{tabular}{c | c c c c c c c c c} 
 \hline
 \thead{ID} & \thead{Age} & \thead{Sex} & \thead{Education} & \thead{Occupation\\before Retirement} & \thead{Past Experience \\ {\it ``I am familiar with VA''}} & \thead{TechPH\\(Enthusiasm)} & \thead{TechPH\\(Anxiety)} & \thead{MDPQ} & \thead{CPQ} \\ [0.5ex] 
 \hline\hline
 P1 & 90 - 95 & F & Doctorate Degree & Social Sciences Researcher & Somewhat Agree (4) & 3.3 & 3.3 & 3.3 & 3.9 \\ 

 P2 & 80 - 85 & F & Professional Degree & Healthcare Worker & Strongly Agree (5) & 4.7 & 2.3 & 4.8 & 4.8 \\

 P3 & 70 - 75 & F & Professional Degree & Administrator & Somewhat Disagree (2) & 3.7 & 4.7 & 2.4 & 4.3 \\

 P4 & 70 - 75 & F & Professional Degree & Administrator & Strongly Agree (5) & 3.3 & 4.7 & 4.3 & 4.4 \\

 P5 & 90 - 95 & M & Professional Degree & Healthcare Worker & Somewhat Agree (4) & 4.3 & 2.0 & 4.8 & 4.9 \\

 P6 & 75 - 80 & M & Some College & Businessman & Somewhat Agree (4) & 4.7 & 3.3 & 4.4 & 4.5 \\

 P7 & 75 - 80 & F & Professional Degree & Librarian & Somewhat Agree (4) & 4.7 & 2.0 & 3.4 & 4.8 \\

 P8 & 75 - 80 & F & Professional Degree & Social Worker & Somewhat Agree (4) & 2.0 & 4.3 & 5.0 & 4.0 \\

 P9 & 70 - 75 & M & Professional Degree & Engineer & Strongly Agree (5) & 5.0 & 2.0 & 4.1 & 4.9 \\

 P10 & 75 - 80 & M & Professional Degree & Consultant & Somewhat Disagree (2) & 4.3 & 3.0 & 1.0 & 4.3 \\
 \hline
\end{tabular}
\caption{Participants' demographic data and the responses from the questionnaires of \emph{TechPH}~\cite{Anderberg2019}, \emph{MDPQ}~\cite{Petrovcic2019}, and \emph{CPQ}~\cite{Boot2015}. All scores are on the scale of $1$ to $5$. Participants with proficient technology use would have high score of VA's past experience, TechPH (Enthusiasm), MDPQ, and CPQ, and a low score of TechPH (Anxiety).}
\vspace{-8mm}
\label{fig::participant}
\end{table*}

Rooted in participant observations and semi-structured interviews, we aim to tackle two key questions:
\textbf{(i)} To what extent are older adults able to independently set up and onboard a standalone voice-only and voice-first VA?
\textbf{(ii)} What are the affordances of the built-in touchscreen for voice-first devices that could help during the device setup procedure?
We formulated our study as a within-subject design, where participants were instructed to set up both the Echo Dot and Echo Show, under the observation of research assistants.
The entire processes were video and audio recorded, with the consent of participants.
Our study procedure consisted of three parts.

\vspace{3px}\noindent
{\bf (1)~Pre-Study Questionnaire.} 
Participants were first invited to complete three well-known questionnaires that aim to evaluate {\bf (i)} older adults' attitudes (\incl~enthusiasm and anxiety) toward general technologies~\cite{Anderberg2019},  {\bf (ii)} older adults' proficiency with smartphones~\cite{Petrovcic2019}, and {\bf (iii)} older adults' proficiency with desktop computers~\cite{Boot2015}.
This aims to offer implications regarding participants' experience and attitude of using general computing devices that might affect their abilities on setting up designated VAs.

\vspace{3px}\noindent
{\bf (2)~Setting Up the Voice Assistants.}
Participants were then instructed to unbox and set up the given Echo Dot ($3$rd Gen, 2018 release) and Echo Show $8$ ($2$nd Gen, 2021 release) with a built-in eight inches touchscreen, as well as connect the devices to their home's WiFi using the official instructions.
While we instructed participants to set up both devices {\it independently}, hints and assistance were provided {\it if and only if} the older adults were having difficulty and actively requested support.
To avoid the impacts of confounding factors of the learning experience, P1 - P5 were instructed to set up Echo Dot, followed by Echo Show, while P6 - P10 were instructed to set up Echo Show, followed by Echo Dot.
Participants were asked to think out loud while completing the given tasks.
Figure~\ref{fig::teaser} illustrates an example scene while setting up the Echo Dot (Fig.~\ref{fig::teaser}a - d) and Echo Show (Fig.~\ref{fig::teaser}e - i).

\vspace{3px}\noindent
{\bf (3) Post-Study Interviews.} 
After setting up both devices, we conducted a semi-structured interview with each participant.
We focused on two key guiding questions: {\bf (i)}{\it``what are the major challenges you experienced while setting up the given devices, and why?''}, and {\bf (ii)} {\it``what potential values do you think that the additional touch display brings, and why?''}~\enspace
The format of the discussions was kept open-ended so that participants could expand on their responses, and raise follow-up discussions as needed.

Our preliminary study reported in this paper focused on analyzing the qualitative data only.
We transcribed the audio recordings and manually edited and/or corrected the transcriptions as needed.
Two research assistants then used deductive and inductive coding approach to analyze participants' comments thematically, with the recorded videos being referred as needed.
\section{Results}\label{sec::results}

Overall, while most participants were able to set up both Echo Dot and Show, different opinions were raised during the interviews. 
In particular, P6 stopped half way while typing the login credentials for Echo Dot; P8 asked for assistance to help her setting up both VAs; and P4 only asked for assistance during the Echo Show setup phase due to unexpected duties unrelated to the study.
We outline our findings spanning across three themes.

\vspace{3px}\noindent
{\bf Device Form Factor Impacts Placing VAs in the Home.}
The form factors of the device, specifically with the size of the device and the power cable and plug, impacted participants’ abilities to easily place the device in the home.
Most participants made interesting comments regarding the potential design considerations of the form factors of both standalone VAs.
Due to the built-in touchscreen, the Echo Show is naturally larger than the Echo Dot, leading to negative impacts on some participants' decisions when it comes to the device placement in their home (see Fig.~\ref{fig::teaser}). 
For example, P2 emphasized the importance of the small form factor and how this guided her decision: {\it ``[...] the Dot is smaller and more inconspicuous. So it’s easier to fit into a smaller space. I like the convenience of that. [...]''}~(P2)
While placing both devices in their home environment, some participants complained about the lack of long power cables to support their ideal placements, and the fact that they had to change their plans. For example P7 commented: {\it ``the cable is too short, so I could only put it right here [participant also tried to point to an awkward place that is hard to reach].''}~(P7)
Similarly, another participant also complained about the problems caused by the large power plug. Despite she had experience setting up Echo Dot, P2 mentioned how: {\it ``one of the challenges of the Amazon plug is that the plugs are very big, so they take much more place than a simple plug, such as the plug for a lamp. So sometimes it is hard to find a suitable place to plug in.''}~(P2)
This led our participants to have to endure awkward and uncomfortable postures while setting up the devices, which in turn might lead to increased health problems, especially for those with mobility impairments (\eg~P1).


\vspace{3px}\noindent
{\bf Challenges with Instruction Manuals.}
Participants overall believed that the instruction documents are important, yet the current design is not senior-friendly.
Participants were instructed to set up the device independently by reading the official instruction document. Some participants raised concerns regarding the design of the user manuals.
P10 summarized well the experience and emphasized the importance of {\it ``test drive''} with older adults users: {\it ``I think that people who are creating these [manuals] are making their own assumptions. They think it’ll be easy for them to do this. But not necessarily...somebody needs to have a test drive with a user like what we’re doing here.''}~(P10)
We also observed that three participants skipped reading the instruction manual for Dot and another three did the same for the Show, and proceeded on setting up devices based on their past experience.
One of the reasons for skipping the manual had to do with small print size, as reported explicitly by P2: {\it ``the print is small. I would have to put on my reading glasses to look at this. [...] Whereas you see on the screen here [On the Echo Show], it’s not a problem. Look how big the type is on the screen. So that’s easy, and you can even see that from a distance''}~(P2).
Others mentioned the setback of text- and image-based instructions and still emphasized the importance of having access to a technical support person: {\it ``I would prefer to go over the information with a person or in some kind of a tutorial [like a video on YouTube], rather than just sitting with the machine and trying to figure it out myself''}~(P8).
P10 proposed an insightful design consideration that, while going against the instruction manual, would take advantage of the interactivity of the built-in touchscreen: {\it ``I’m thinking about not using a piece of paper here [...], just doing the whole thing on the screen''}~(P10).
Finally, some participants reported how the instructions were missing critical details. 
For example, P3 outlined how {\it ``I tend to ignore instructions because they’re usually very poorly written''}, and P8 confirmed that {\it ``it’s knowing what things mean, that’s the hardest!''}

\vspace{3px}\noindent{\bf Affordances of the Built--In Touchscreen.}
Overall, participants strongly preferred the experience of setting up VAs with the additional touchscreen since the screen allowed for more visual cues and intuitive data entry.
Through participant observations, we found that older adults used different strategies to type information on mobile devices, such as using a stylus (P6), an external keyboard (P9), or a tablet with a bigger display (P10).
Overall, all participants strongly preferred the experience introduced by the additional touchscreen. P6 acknowledged that the display simplifies the process of setting up the device: {\it ``I think that [the device setup on Show] is easier because of the screen''}. 
Participants also explained how the additional touch screen could be helpful regarding the device setup experience.
For example, P2 and P3 outlined how ease of typing is an important benefit: {\it ``inputting the data is the most helpful! because the screen was bigger than my phone.''} (P3), {\it ``the underscore sign is a little bit hard to be found on this phone.''} (P2). 
P7 stressed how she preferred the interaction with the keyboard on the display of Echo Show: {\it ``it’s easier to see and easier to handle with the keyboard on the display [...] I’m used to my phone. But this [the keyboard on the Show] has bigger buttons''} (P7).
P3 emphasized the issues of the well–-known ``fat finger'' problems~\cite{Siek2005}: {\it ``[with my phone] I make a lot of mistakes, because it’s small. And I miss entering information with my fat fingers.''}~(P3)
Participants also preferred the visual feedback created by the touchscreen, compared to the prompts on the smartphone app. This might be due to the immediate feedback given by Show, and the consequent reduced demands on users’ working memory.
Example testimonies include: 
{\it ``the setup was easier on the Show. Because we could actually see what we were doing. Whereas [with Dot] you’re only hearing it and seeing it on the phone.''} (P2),
{\it``[Echo Show] gives me the directions right on the screen, then it would be easier than me looking at my phone and transferring the information mentally.''} (P4),
and 
{\it``I prefer visual rather than just auditory in general.''} (P8).
The lack of direct and \insitu visual guidance while setting up the Echo Dot (as mentioned by P4) could be one possible cause of three participants failing to find the setup buttons without hints. 
Since for Echo Dot, part of the instructions were on the Alexa phone app, when instructed to {\it ``touch''} the setup button, some participants (\eg~P1) incorrectly considered the button icon on the phone as the target, while others (P4) made incorrect attempts to interact with the mute button on the Dot. 
Some participants' think-out-loud comments outline well these observations: {\it ``I had assumed the button to push was the one on the top rather than the one on the side. [...] [The system should] tell me which button to push more precisely''}~(P4).

\section{Lessons Learned and Future Work}\label{sec::discussion}
In conclusion, we summarize the three main lessons learnt that will inform our future research that aims to design and create senior-friendly VAs for a wide array of general uses:

\vspace{3px}\noindent
{\bf Recommendations of Device Placements Should be Offered.}
We found that nearly all participants have difficulties on finding a suitable position with reachable distance to the power outlet in their home to place their devices, due to the {``unexpected''} form factors.
Future design could possibly include more interactive recommendations where the devices could potentially be placed and be integrated as part of older adults' home.

\vspace{3px}\noindent
{\bf Instructions Should be Readable, Interactive, and Consider Older Adults' Needs.}
While instruction manuals are designed to help users better and easily set up and onboarding with the devices, we showed that older adults expected the instruction documents to have larger font size and be more interactive.
The written content in the instructions should also be validated by {``pilot''} older adult users, as outlined by P10.
Future research might investigate how to design a usable instruction documents for VAs, and how to create an interactive and usable instruction guide with voice user interfaces and touchscreen (for scree-based VAs).
Additional future work might also explore how to design a usable instruction experience for older adults with different abilities (Table.~\ref{fig::participant}). 

\vspace{3px}\noindent
{\bf Built-In Touchscreen Should be Leveraged while Setting Up and Onboarding with Voice-First Screen-Based VAs.}.
Participants have raised many positive comments and opportunities of touchscreen as evidenced in \S\ref{sec::results}.
Future research might consider novel ways and methods to better integrate the touchscreen as part of {``interactive''} setup instructions, which could faithfully replace the often required technical support person.

\begin{acks}
This work is part of project VOLI\footnote{Project VOLI $@$UC San Diego: \href{http://voli.ucsd.edu}{http://voli.ucsd.edu} [Last Accessed on May 12, 2023].} and was supported by NIH/NIA under grant R56AG067393.
Co-author Michael Hogarth has an equity interest in LifeLink Inc. and also serves on the company’s Scientific Advisory Board. The terms of this arrangement have been reviewed and approved by the UC San Diego in accordance with its conflict of interest policies.
We appreciate Mary Draper and other residents from \mbox{the Vi at La Jolla} for helping with participant recruitment.
\end{acks}

\balance
\bibliographystyle{ACM-Reference-Format}
\bibliography{reference}


\begin{thebibliography}{25}


\ifx \showCODEN    \undefined \def \showCODEN     #1{\unskip}     \fi
\ifx \showDOI      \undefined \def \showDOI       #1{#1}\fi
\ifx \showISBNx    \undefined \def \showISBNx     #1{\unskip}     \fi
\ifx \showISBNxiii \undefined \def \showISBNxiii  #1{\unskip}     \fi
\ifx \showISSN     \undefined \def \showISSN      #1{\unskip}     \fi
\ifx \showLCCN     \undefined \def \showLCCN      #1{\unskip}     \fi
\ifx \shownote     \undefined \def \shownote      #1{#1}          \fi
\ifx \showarticletitle \undefined \def \showarticletitle #1{#1}   \fi
\ifx \showURL      \undefined \def \showURL       {\relax}        \fi
\providecommand\bibfield[2]{#2}
\providecommand\bibinfo[2]{#2}
\providecommand\natexlab[1]{#1}
\providecommand\showeprint[2][]{arXiv:#2}

\bibitem[Anderberg et~al\mbox{.}(2019)]%
        {Anderberg2019}
\bibfield{author}{\bibinfo{person}{Peter Anderberg}, \bibinfo{person}{Shahryar
  Eivazzadeh}, \bibinfo{person}{Johan~Sanmartin Berglund}, {et~al\mbox{.}}}
  \bibinfo{year}{2019}\natexlab{}.
\newblock \showarticletitle{A {N}ovel {I}nstrument for {M}easuring {O}lder
  {P}eople’s {A}ttitudes {T}oward {T}echnology (TechPH): {D}evelopment and
  {V}alidation}.
\newblock \bibinfo{journal}{\emph{{J}ournal of {M}edical Internet {R}esearch}}
  \bibinfo{volume}{21}, \bibinfo{number}{5} (\bibinfo{year}{2019}).
\newblock
\urldef\tempurl%
\url{https://doi.org/10.2196/13951}
\showDOI{\tempurl}


\bibitem[Bajorek(2018)]%
        {Bajorek2018}
\bibfield{author}{\bibinfo{person}{Joan~Palmiter Bajorek}.}
  \bibinfo{year}{2018}\natexlab{}.
\newblock \bibinfo{booktitle}{\emph{Voice First Versus the Multimodal User
  Interfaces of the Future}}.
\newblock
\urldef\tempurl%
\url{https://www.uxmatters.com/mt/archives/2018/10/voice-first-versus-the-multimodal-user-interfaces-of-the-future.php}
\showURL{%
\tempurl}


\bibitem[Boldy et~al\mbox{.}(2011)]%
        {Boldy2011}
\bibfield{author}{\bibinfo{person}{Duncan Boldy}, \bibinfo{person}{Linda
  Grenade}, \bibinfo{person}{Gill Lewin}, \bibinfo{person}{Elizabeth Karol},
  {and} \bibinfo{person}{Elissa Burton}.} \bibinfo{year}{2011}\natexlab{}.
\newblock \showarticletitle{Older people's decisions regarding ‘ageing in
  place’: A Western Australian case study}.
\newblock \bibinfo{journal}{\emph{Australasian Journal on Ageing}}
  \bibinfo{volume}{30}, \bibinfo{number}{3} (\bibinfo{year}{2011}),
  \bibinfo{pages}{136--142}.
\newblock
\urldef\tempurl%
\url{https://doi.org/10.1111/j.1741-6612.2010.00469.x}
\showDOI{\tempurl}


\bibitem[Boot et~al\mbox{.}(2015)]%
        {Boot2015}
\bibfield{author}{\bibinfo{person}{Walter~R Boot}, \bibinfo{person}{Neil
  Charness}, \bibinfo{person}{Sara~J Czaja}, \bibinfo{person}{Joseph Sharit},
  \bibinfo{person}{Wendy~A Rogers}, \bibinfo{person}{Arthur~D Fisk},
  \bibinfo{person}{Tracy Mitzner}, \bibinfo{person}{Chin~Chin Lee}, {and}
  \bibinfo{person}{Sankaran Nair}.} \bibinfo{year}{2015}\natexlab{}.
\newblock \showarticletitle{{C}omputer {P}roficiency {Q}uestionnaire:
  {A}ssessing {L}ow and {H}igh {C}omputer {P}roficient {S}eniors}.
\newblock \bibinfo{journal}{\emph{The Gerontologist}} \bibinfo{volume}{55},
  \bibinfo{number}{3} (\bibinfo{year}{2015}), \bibinfo{pages}{404--411}.
\newblock
\urldef\tempurl%
\url{https://doi.org/10.1093/geront/gnt117}
\showDOI{\tempurl}


\bibitem[Charles et~al\mbox{.}(2021)]%
        {Charles2021}
\bibfield{author}{\bibinfo{person}{Kemeberley Charles}, \bibinfo{person}{Chen
  Chen}, \bibinfo{person}{Janet~G. Johnson}, \bibinfo{person}{Alice Lee},
  \bibinfo{person}{Ella~T. Lifset}, \bibinfo{person}{Michael Hogarth},
  \bibinfo{person}{Nadir Weibel}, \bibinfo{person}{Emilia Farcas}, {and}
  \bibinfo{person}{Alison~A. Moore}.} \bibinfo{year}{2021}\natexlab{}.
\newblock \showarticletitle{{H}ow {M}ight an {I}ntelligent {V}oice {A}ssistant
  {A}ddress {O}lder {A}dults' {H}ealth-related {N}eeds?}. In
  \bibinfo{booktitle}{\emph{{J}ournal of the {A}merican {G}eriatrics
  {S}ociety}}, Vol.~\bibinfo{volume}{69}. Wiley 111 River St, Hoboken
  07030-5774, NJ, USA, \bibinfo{pages}{S243--S244}.
\newblock


\bibitem[Chen et~al\mbox{.}(2021a)]%
        {Chen2021assets}
\bibfield{author}{\bibinfo{person}{Chen Chen}, \bibinfo{person}{Janet~G.
  Johnson}, \bibinfo{person}{Charles Kemeberley}, \bibinfo{person}{Alice Lee},
  \bibinfo{person}{Ella~T. Lifset}, \bibinfo{person}{Michael Hogarth},
  \bibinfo{person}{Alison~A. Moore}, \bibinfo{person}{Emilia Farcas}, {and}
  \bibinfo{person}{Nadir Weibel}.} \bibinfo{year}{2021}\natexlab{a}.
\newblock \showarticletitle{{U}nderstanding {B}arriers and {D}esign
  {O}pportunities to {I}mprove {H}ealthcare and {QOL} for {O}lder {A}dults
  through {V}oice {A}ssistants}. In \bibinfo{booktitle}{\emph{The 22nd
  International ACM SIGACCESS Conference on Computers and Accessibility}}
  (Virtual Event, USA) \emph{(\bibinfo{series}{ASSETS' 21})}.
  \bibinfo{publisher}{Association for Computing Machinery}.
\newblock
\urldef\tempurl%
\url{https://doi.org/10.1145/3441852.3471218}
\showDOI{\tempurl}


\bibitem[Chen et~al\mbox{.}(2021b)]%
        {Chen2021}
\bibfield{author}{\bibinfo{person}{Chen Chen}, \bibinfo{person}{Khalil Mrini},
  \bibinfo{person}{Kemeberly Charles}, \bibinfo{person}{Ella Lifset},
  \bibinfo{person}{Michael Hogarth}, \bibinfo{person}{Alison Moore},
  \bibinfo{person}{Nadir Weibel}, {and} \bibinfo{person}{Emilia Farcas}.}
  \bibinfo{year}{2021}\natexlab{b}.
\newblock \showarticletitle{Toward a Unified Metadata Schema for Ecological
  Momentary Assessment with Voice-First Virtual Assistants}. In
  \bibinfo{booktitle}{\emph{CUI 2021 - 3rd Conference on Conversational User
  Interfaces}} (Bilbao (online), Spain) \emph{(\bibinfo{series}{CUI '21})}.
  \bibinfo{publisher}{Association for Computing Machinery},
  \bibinfo{address}{New York, NY, USA}, Article \bibinfo{articleno}{31},
  \bibinfo{numpages}{6}~pages.
\newblock
\showISBNx{9781450389983}
\urldef\tempurl%
\url{https://doi.org/10.1145/3469595.3469626}
\showDOI{\tempurl}


\bibitem[Choi et~al\mbox{.}(2018)]%
        {Choi2018}
\bibfield{author}{\bibinfo{person}{Y Choi}, \bibinfo{person}{G Demiris}, {and}
  \bibinfo{person}{H Thompson}.} \bibinfo{year}{2018}\natexlab{}.
\newblock \showarticletitle{Feasibility of {S}mart {S}peaker {U}se to {S}upport
  {A}ging in {P}lace}.
\newblock \bibinfo{journal}{\emph{Innovation in aging}} \bibinfo{volume}{2},
  \bibinfo{number}{suppl\_1} (\bibinfo{year}{2018}), \bibinfo{pages}{560--560}.
\newblock
\urldef\tempurl%
\url{https://doi.org/10.1093/geroni/igy023.2073}
\showDOI{\tempurl}


\bibitem[Czaja(1997)]%
        {Marting1997}
\bibfield{author}{\bibinfo{person}{Sara~J. Czaja}.}
  \bibinfo{year}{1997}\natexlab{}.
\newblock \showarticletitle{Computer Technology and the Older Adult}.
\newblock In \bibinfo{booktitle}{\emph{Handbook of Human-Computer Interaction
  (Second Edition)} (\bibinfo{edition}{second edition} ed.)},
  \bibfield{editor}{\bibinfo{person}{Marting~G. Helander},
  \bibinfo{person}{Thomas~K. Landauer}, {and} \bibinfo{person}{Prasad~V.
  Prabhu}} (Eds.). \bibinfo{publisher}{North-Holland},
  \bibinfo{address}{Amsterdam}, \bibinfo{pages}{797--812}.
\newblock
\showISBNx{978-0-444-81862-1}
\urldef\tempurl%
\url{https://doi.org/10.1016/B978-044481862-1.50100-X}
\showDOI{\tempurl}


\bibitem[Eckert et~al\mbox{.}(2004)]%
        {Eckert2004}
\bibfield{author}{\bibinfo{person}{J~Kevin Eckert}, \bibinfo{person}{Leslie~A
  Morgan}, {and} \bibinfo{person}{Namratha Swamy}.}
  \bibinfo{year}{2004}\natexlab{}.
\newblock \showarticletitle{{P}references for {R}eceipt of {C}are {A}mong
  {C}ommunity-dwelling {A}dults}.
\newblock \bibinfo{journal}{\emph{Journal of aging \& social policy}}
  \bibinfo{volume}{16}, \bibinfo{number}{2} (\bibinfo{year}{2004}),
  \bibinfo{pages}{49--65}.
\newblock
\urldef\tempurl%
\url{https://doi.org/10.1300/J031v16n02_04}
\showDOI{\tempurl}


\bibitem[Han et~al\mbox{.}(2022)]%
        {Han2022}
\bibfield{author}{\bibinfo{person}{Yichen Han}, \bibinfo{person}{Christopher~Bo
  Han}, \bibinfo{person}{Chen Chen}, \bibinfo{person}{Peng~Wei Lee},
  \bibinfo{person}{Michael Hogarth}, \bibinfo{person}{Alison~A. Moore},
  \bibinfo{person}{Nadir Weibel}, {and} \bibinfo{person}{Emilia Farcas}.}
  \bibinfo{year}{2022}\natexlab{}.
\newblock \showarticletitle{Towards Visualization of Time–Series Ecological
  Momentary Assessment (EMA) Data on Standalone Voice–First Virtual
  Assistants}. In \bibinfo{booktitle}{\emph{Proceedings of the 24th
  International ACM SIGACCESS Conference on Computers and Accessibility}}
  (Athens, Greece) \emph{(\bibinfo{series}{ASSETS '22})}.
  \bibinfo{publisher}{Association for Computing Machinery},
  \bibinfo{address}{New York, NY, USA}, Article \bibinfo{articleno}{60},
  \bibinfo{numpages}{4}~pages.
\newblock
\showISBNx{9781450392587}
\urldef\tempurl%
\url{https://doi.org/10.1145/3517428.3550398}
\showDOI{\tempurl}


\bibitem[Heinz et~al\mbox{.}(2013)]%
        {Heinz2013}
\bibfield{author}{\bibinfo{person}{Melinda Heinz}, \bibinfo{person}{Peter
  Martin}, \bibinfo{person}{Jennifer~A Margrett}, \bibinfo{person}{Mary
  Yearns}, \bibinfo{person}{Warren Franke}, \bibinfo{person}{Hen-I Yang},
  \bibinfo{person}{Johnny Wong}, {and} \bibinfo{person}{Carl~K Chang}.}
  \bibinfo{year}{2013}\natexlab{}.
\newblock \showarticletitle{{P}erceptions of {T}echnology {A}mong {O}lder
  {A}dults}.
\newblock \bibinfo{journal}{\emph{Journal of gerontological nursing}}
  \bibinfo{volume}{39}, \bibinfo{number}{1} (\bibinfo{year}{2013}),
  \bibinfo{pages}{42--51}.
\newblock
\urldef\tempurl%
\url{https://doi.org/10.3928/00989134-20121204-04}
\showDOI{\tempurl}


\bibitem[Jansons et~al\mbox{.}(2022)]%
        {Jansons2022}
\bibfield{author}{\bibinfo{person}{Paul Jansons}, \bibinfo{person}{Jackson
  Fyfe}, \bibinfo{person}{Jack Dalla~Via}, \bibinfo{person}{Robin~M Daly},
  \bibinfo{person}{Eugene Gvozdenko}, {and} \bibinfo{person}{David Scott}.}
  \bibinfo{year}{2022}\natexlab{}.
\newblock \showarticletitle{Barriers and Enablers for Older Adults
  Participating in A Home-Based Pragmatic Exercise Program Delivered and
  Monitored by Amazon Alexa: A Qualitative Study}.
\newblock  (\bibinfo{year}{2022}).
\newblock
\urldef\tempurl%
\url{https://doi.org/10.1186/s12877-022-02963-2}
\showDOI{\tempurl}


\bibitem[Kim and Choudhury(2021)]%
        {Sunyoung2021}
\bibfield{author}{\bibinfo{person}{Sunyoung Kim} {and}
  \bibinfo{person}{Abhishek Choudhury}.} \bibinfo{year}{2021}\natexlab{}.
\newblock \showarticletitle{Exploring {O}lder {A}dults’ {P}erception and
  {U}se of {S}mart {S}peaker-based {V}oice {A}ssistants: A {L}ongitudinal
  {S}tudy}.
\newblock \bibinfo{journal}{\emph{Computers in Human Behavior}}
  \bibinfo{volume}{124} (\bibinfo{year}{2021}), \bibinfo{pages}{106914}.
\newblock
\showISSN{0747-5632}
\urldef\tempurl%
\url{https://doi.org/10.1016/j.chb.2021.106914}
\showDOI{\tempurl}


\bibitem[Kinsella(2020)]%
        {Kinsella2020}
\bibfield{author}{\bibinfo{person}{Bret Kinsella}.}
  \bibinfo{year}{2020}\natexlab{}.
\newblock \bibinfo{booktitle}{\emph{Amazon Smart Speaker Market Share Falls to
  53\% in 2019 with Google The Biggest Beneficiary Rising to 31\%, Sonos Also
  Moves Up}}.
\newblock
\urldef\tempurl%
\url{https://voicebot.ai/2020/04/28/amazon-smart-speaker-market-share-falls-to-53-in-2019-with-google-the-biggest-beneficiary-rising-to-31-sonos-also-moves-up/}
\showURL{%
\tempurl}


\bibitem[Lifset et~al\mbox{.}(2020)]%
        {Lifset2020}
\bibfield{author}{\bibinfo{person}{Ella~T. Lifset}, \bibinfo{person}{Kemeberly
  Charles}, \bibinfo{person}{Emilia Farcas}, \bibinfo{person}{Nadir Weibel},
  \bibinfo{person}{Michael Hogarth}, \bibinfo{person}{Chen Chen},
  \bibinfo{person}{Janet Johnson}, {and} \bibinfo{person}{Alison Moore}.}
  \bibinfo{year}{2020}\natexlab{}.
\newblock \showarticletitle{Can an Intelligent Virtual Assistant (IVA) Meet
  Older Adult Health-Related Needs in the Context of a Geriatric 5Ms
  Framework?}. In \bibinfo{booktitle}{\emph{Journal OF the American Geriatrics
  Society}}, Vol.~\bibinfo{volume}{70}. Wiley 111 River St., Hoboken
  07030-5774, NJ, USA, \bibinfo{pages}{S245--S246}.
\newblock


\bibitem[Matsumoto et~al\mbox{.}(2016)]%
        {Matsumoto2016}
\bibfield{author}{\bibinfo{person}{Hiroshige Matsumoto},
  \bibinfo{person}{Takashi Naruse}, \bibinfo{person}{Mahiro Sakai}, {and}
  \bibinfo{person}{Satoko Nagata}.} \bibinfo{year}{2016}\natexlab{}.
\newblock \showarticletitle{Who Prefers to Age in Place? Cross-sectional Survey
  of Middle-aged People in Japan}.
\newblock \bibinfo{journal}{\emph{Geriatrics \& Gerontology International}}
  \bibinfo{volume}{16}, \bibinfo{number}{5} (\bibinfo{year}{2016}),
  \bibinfo{pages}{631--637}.
\newblock
\urldef\tempurl%
\url{https://doi.org/10.1111/ggi.12503}
\showDOI{\tempurl}


\bibitem[Mrini et~al\mbox{.}(2021)]%
        {Mrini2021}
\bibfield{author}{\bibinfo{person}{Khalil Mrini}, \bibinfo{person}{Chen Chen},
  \bibinfo{person}{Ndapa Nakashole}, \bibinfo{person}{Nadir Weibel}, {and}
  \bibinfo{person}{Emilia Farcas}.} \bibinfo{year}{2021}\natexlab{}.
\newblock \showarticletitle{Medical Question Understanding and Answering for
  Older Adults}.
\newblock  (\bibinfo{year}{2021}).
\newblock


\bibitem[Nallam et~al\mbox{.}(2020)]%
        {Nallam2020}
\bibfield{author}{\bibinfo{person}{Phani Nallam}, \bibinfo{person}{Siddhant
  Bhandari}, \bibinfo{person}{Jamie Sanders}, {and} \bibinfo{person}{Aqueasha
  Martin-Hammond}.} \bibinfo{year}{2020}\natexlab{}.
\newblock \showarticletitle{A {Q}uestion of {A}ccess: Exploring the {P}erceived
  {B}enefits and {B}arriers of {I}ntelligent {V}oice {A}ssistants for
  {I}mproving {A}ccess to {C}onsumer {H}ealth {R}esources {A}mong {L}ow-income
  {O}lder {A}dults}.
\newblock \bibinfo{journal}{\emph{Gerontology and Geriatric Medicine}}
  \bibinfo{volume}{6} (\bibinfo{year}{2020}),
  \bibinfo{pages}{2333721420985975}.
\newblock
\urldef\tempurl%
\url{https://doi.org/10.1177/2333721420985975}
\showDOI{\tempurl}


\bibitem[Pang et~al\mbox{.}(2021)]%
        {Pang2021}
\bibfield{author}{\bibinfo{person}{Carolyn Pang}, \bibinfo{person}{Zhiqin
  Collin~Wang}, \bibinfo{person}{Joanna McGrenere}, \bibinfo{person}{Rock
  Leung}, \bibinfo{person}{Jiamin Dai}, {and} \bibinfo{person}{Karyn Moffatt}.}
  \bibinfo{year}{2021}\natexlab{}.
\newblock \showarticletitle{Technology Adoption and Learning Preferences for
  Older Adults: Evolving Perceptions, Ongoing Challenges, and Emerging Design
  Opportunities}. In \bibinfo{booktitle}{\emph{Proceedings of the 2021 CHI
  Conference on Human Factors in Computing Systems}} (Yokohama, Japan)
  \emph{(\bibinfo{series}{CHI '21})}. \bibinfo{publisher}{Association for
  Computing Machinery}, \bibinfo{address}{New York, NY, USA}, Article
  \bibinfo{articleno}{490}, \bibinfo{numpages}{13}~pages.
\newblock
\showISBNx{9781450380966}
\urldef\tempurl%
\url{https://doi.org/10.1145/3411764.3445702}
\showDOI{\tempurl}


\bibitem[Petrovčič et~al\mbox{.}(2019)]%
        {Petrovcic2019}
\bibfield{author}{\bibinfo{person}{Andraž Petrovčič},
  \bibinfo{person}{Walter~R. Boot}, \bibinfo{person}{Tomaž Burnik}, {and}
  \bibinfo{person}{Vesna Vesna~Dolničar}.} \bibinfo{year}{2019}\natexlab{}.
\newblock \showarticletitle{Improving the Measurement of Older Adults’ Mobile
  Device Proficiency: Results and Implications from a Study of Older Adult
  Smartphone Users}.
\newblock \bibinfo{journal}{\emph{IEEE Access}}  \bibinfo{volume}{7}
  (\bibinfo{year}{2019}), \bibinfo{pages}{150412--150422}.
\newblock
\urldef\tempurl%
\url{https://doi.org/10.1109/ACCESS.2019.2947765}
\showDOI{\tempurl}


\bibitem[Shade et~al\mbox{.}(2021)]%
        {Shade2021}
\bibfield{author}{\bibinfo{person}{Marcia Shade}, \bibinfo{person}{Kyle
  Rector}, \bibinfo{person}{Kevin Kupzyk}, {et~al\mbox{.}}}
  \bibinfo{year}{2021}\natexlab{}.
\newblock \showarticletitle{Voice Assistant Reminders and the Latency of
  Scheduled Medication Use in Older Adults With Pain: Descriptive Feasibility
  Study}.
\newblock \bibinfo{journal}{\emph{JMIR Formative Research}}
  \bibinfo{volume}{5}, \bibinfo{number}{9} (\bibinfo{year}{2021}),
  \bibinfo{pages}{e26361}.
\newblock
\urldef\tempurl%
\url{https://doi.org/10.2196/26361}
\showDOI{\tempurl}


\bibitem[Siek et~al\mbox{.}(2005)]%
        {Siek2005}
\bibfield{author}{\bibinfo{person}{Katie~A Siek}, \bibinfo{person}{Yvonne
  Rogers}, {and} \bibinfo{person}{Kay~H Connelly}.}
  \bibinfo{year}{2005}\natexlab{}.
\newblock \showarticletitle{Fat Finger Worries: How Older and Younger Users
  Physically Interact with PDAs}. In \bibinfo{booktitle}{\emph{IFIP Conference
  on Human-Computer Interaction}}. Springer, \bibinfo{pages}{267--280}.
\newblock
\urldef\tempurl%
\url{https://doi.org/10.1007/11555261_24}
\showDOI{\tempurl}


\bibitem[Trajkova and Martin-Hammond(2020)]%
        {Trajkova2020}
\bibfield{author}{\bibinfo{person}{Milka Trajkova} {and}
  \bibinfo{person}{Aqueasha Martin-Hammond}.} \bibinfo{year}{2020}\natexlab{}.
\newblock \showarticletitle{``Alexa is a Toy'': Exploring Older Adults’
  Reasons for Using, Limiting, and Abandoning Echo}. In
  \bibinfo{booktitle}{\emph{Proceedings of the 2020 CHI Conference on Human
  Factors in Computing Systems}} (Honolulu, HI, USA)
  \emph{(\bibinfo{series}{CHI ’20})}. \bibinfo{publisher}{Association for
  Computing Machinery}, \bibinfo{address}{New York, NY, USA},
  \bibinfo{pages}{1–13}.
\newblock
\showISBNx{9781450367080}
\urldef\tempurl%
\url{https://doi.org/10.1145/3313831.3376760}
\showDOI{\tempurl}


\bibitem[Whitenton(2017)]%
        {Whitenton2017}
\bibfield{author}{\bibinfo{person}{Kathryn Whitenton}.}
  \bibinfo{year}{2017}\natexlab{}.
\newblock \bibinfo{booktitle}{\emph{Voice First: The Future of Interaction?}}
\newblock
\urldef\tempurl%
\url{https://www.nngroup.com/articles/voice-first}
\showURL{%
\tempurl}


\end{thebibliography}

\end{document}